\documentclass[12pt,preprint]{aastex}
\usepackage{epstopdf}
\usepackage{rotating}

\shorttitle{Superluminal Spot Pairs: Sweeping}
\shortauthors{Nemiroff}

\begin{document}
\title{Superluminal Spot Pair Events in Astronomical Settings: \\ Sweeping Beams}
\author{Robert J. Nemiroff}
\affil{Department of Physics, Michigan Technological University, Houghton, MI 49931}
\email{nemiroff@mtu.edu}

\begin{abstract}
Sweeping beams of light can cast spots moving with superluminal speeds across scattering surfaces. Such faster-than-light speeds are well-known phenomena that do not violate special relativity. It is shown here that under certain circumstances, superluminal spot pair creation and annihilation events can occur that provide unique information to observers. These spot pair events are {\it not} particle pair events -- they are the sudden creation or annihilation of a pair of relatively illuminated spots on a scattering surface. Real spot pair illumination events occur unambiguously on the scattering surface when spot speeds diverge, while virtual spot pair events are observer dependent and perceived only when real spot radial speeds cross the speed of light. Specifically, a virtual spot pair creation event will be observed when a real spot's speed toward the observer drops below $c$, while a virtual spot pair annihilation event will be observed when a real spot's radial speed away from the observer rises above $c$. Superluminal spot pair events might be found angularly, photometrically, or polarimetrically, and might carry useful geometry or distance information. Two example scenarios are briefly considered. The first is a beam swept across a scattering spherical object, exemplified by spots of light moving across Earth's Moon and pulsar companions. The second is a beam swept across a scattering planar wall or linear filament, exemplified by spots of light moving across variable nebulae including Hubble's Variable Nebula. In local cases where the sweeping beam can be controlled and repeated, a three-dimensional map of a target object can be constructed. Used tomographically, this imaging technique is fundamentally different from lens photography, radar, and conventional lidar. \end{abstract}
\keywords{ISM: kinematics and dynamics -- relativistic processes -- scattering}

\section{Introduction}

Although light and all radiations are constrained to travel at the local speed of light $c$ or below, such a limit does not apply to spots of light and boundary shadows that sweep across common scattering surfaces. Such locally superluminal motions do not violate special relativity and cannot be used as a means of local communication (see, for example, Griffiths 1994, or Steane 2012). The observation of superluminal spot pair events, however, may signal to a distant observer that a specific geometry is present. 

Superluminal motions for images, spots, and projected boundaries are not new to physics or astrophysics. It is well known that objects, in particular blobs emanating from quasars, moving less than but close to $c$ toward the observer can appear to separate superluminally (Blandford, McKee, and Rees 1977). Superluminal spot pair creation from a sweeping beam was discussed previously in the context of quasars and AGN by Cavaliere et al. (1971) and mentioned more recently by Baune (2009). Another system where images may appear to exceed the speed of light is gravitational lensing, in particular when images appear near the Einstein ring (Nemiroff 1993) or the source approaches or crosses a lens caustic. 

In this paper the general case of how sweeping beams can create superluminal spots on scattering surfaces will be analyzed, with emphasis on observable spot pair events. Section 2 will discuss general kinematic aspects of superluminal spot pair motion, while in the next two sections, two geometric scenarios will be specifically considered. Section 3 will give an analysis of a superluminal beam sweeping across a distant sphere, while Section 4 will analyze a sweeping beam scattering off of a planar wall or linear filament. In each case the focus will be on the occurrence of spot pairs with a following discussion of astronomical settings where such pairs might be found. Section 5 will give a discussion that includes the possibility that superluminal spot pair events might be used to create three dimensional images of local objects, and conclusions.

\section{General Kinematics}

Sweeping beams occur in many astronomical settings including eclipses, precessing jets, rotating pulsars, expanding AGN clouds, and dust clouds moving in front of stars. The speed of reflected spots and shadows from sweeping beams can be arbitrarily high. For example, although light takes about 0.0116 seconds to cross the Moon, a person can sweep a laser pointer across the Moon's surface in less time. How such superluminal motions appear to an observer can be counter-intuitive. 

Reflecting surfaces discussed here are assumed {\it not} to be mirrors but appear dull and so have reflection properties common for arbitrary scattering surfaces in the universe. Such surfaces will typically be assumed to scatter incoming light in accordance with Lambert's law so that no direction is preferred, although precise Lambertian adherence is not essential to the logic of the analyses. Therefore, to be clear, an observer is assumed able to detect a spot on a scattering surface even if they are not at the angle of exact mirror symmetry. 

It will be assumed here that light travels only in straight lines. Therefore locations on a scattering surface become illuminated only by the exact number of times that a beam source points directly at them. For example, no single place on a scattering surface will be illuminated -- nor will appear to any observer be illuminated -- twice by a beam that sweeps past only once. 

The example case most commonly assumed here will be for a small bright spot moving across a large and opaque scattering surface. Although these zero-dimensional spots could be a one-dimensional boundary -- for example the divider of a truncated plane of light advancing along a dark body -- it is typically assumed here that a small spot is created by a localized beam. Common visualizations of this include bright spots resulting from the scattering of a flashlight beam or a laser. The formalism and results presented here will usually work equally well for extended light boundaries or dark spots -- for example shadows.

Two types of spots will be described here. ``Real" spots are actual locations on the scattering surface illuminated by the sweeping beam. An observer situated on the scattering surface could detect real spots. ``Virtual" spots, contrastingly, are spot locations perceived as illuminated by a distant observer. Virtual spots are observer dependent and could be considered images of real spots. 

\begin{figure}[h]
\includegraphics[width=18cm]{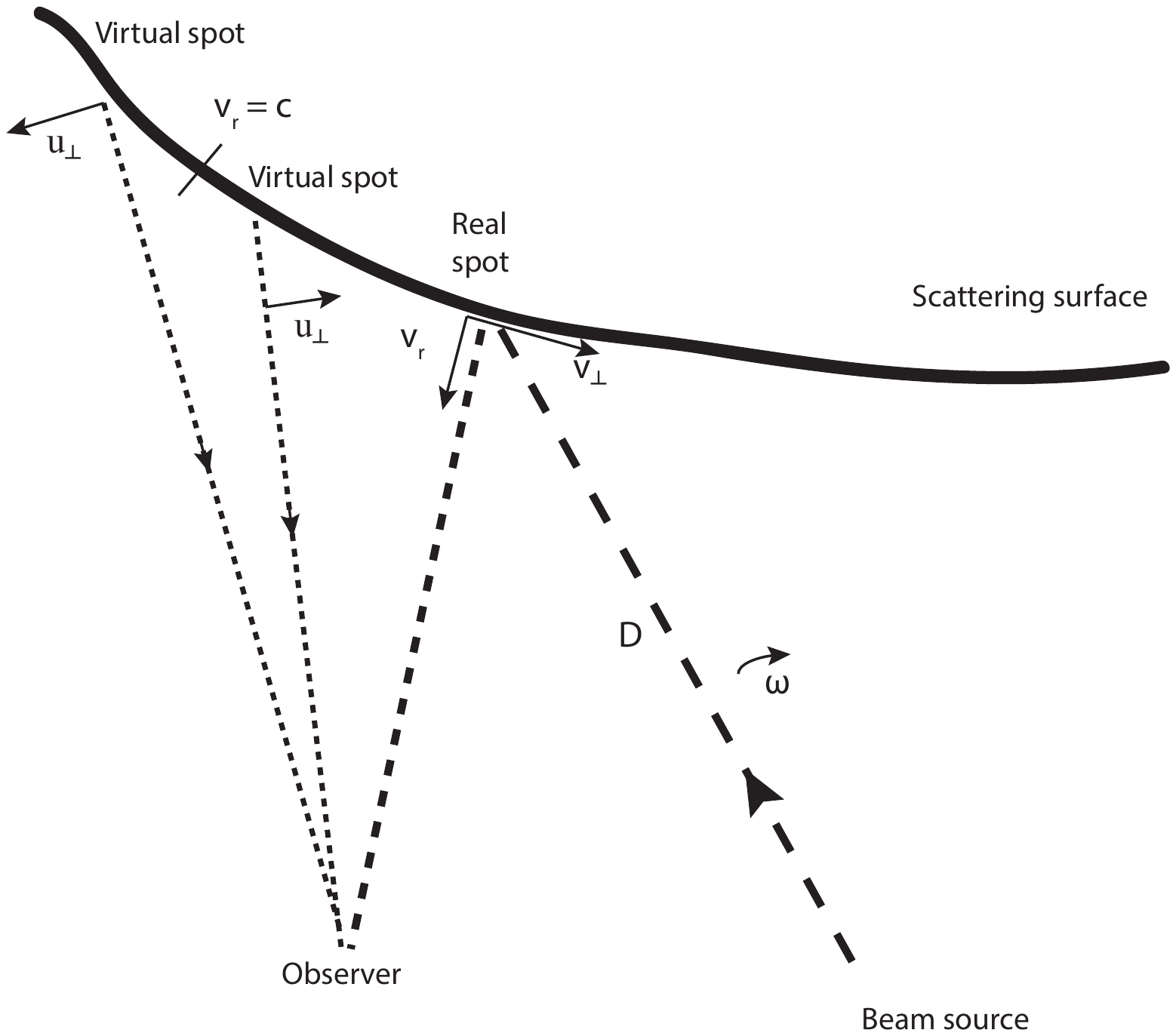}
\caption{A diagram illustrating the three velocities used in the analysis: $w_\perp$, $v$, and $u_\perp$. Here $w_\perp =\omega D$ is the transverse speed of the sweeping beam at the distance $D$ of the scattering surface. Next $v$ is the speed of the real spot on the scattering surface with a component $v_r$ toward the observer and $v_\perp$ perpendicular to the observer. Last $u_\perp$ is the transverse speed of a virtual spot on the scattering surface, as perceived by the observer. The diagram indicates that although one real spot exists at this hypothetical time, two virtual spots appear to the observer on either side of the $v_r = c$ location.}
\label{wvu}
\end{figure}

Three types of velocities will be referred to here. The first velocity will be designated $w_\perp = \omega D$ where $\omega$ is the angular speed of the sweeping beam and $D$ is the distance between the source and the scattering surface. This speed may not describe any actual spot motion and corresponds to the theoretical spot speed on a spherical shell of radius $D$ centered on the beam source. The second velocity will be designated $v$ and refers to the real velocity of the real spot on the reflector. It is useful to break up $v$ into two components. The component radially toward the observer will be designated $v_r$ and the component perpendicular to the observer will be designated $v_\perp$, such that $v^2 = v_r^2 + v_\perp^2$. For simplicity in the cases described herein, a positive value of $v_r$ will be attributed to radial motion toward the observer, while a negative $v_r$ will describe radial motion away from the observer. The third velocity will be designated $u_\perp$ and will refer to the transverse speed of a virtual spot perceived by the observer on the scatterer. Due to finite light travel times between the scatterer and the observer, in general, $u_\perp \ne v_\perp \ne w_\perp$. All three velocities are depicted in Figure \ref{wvu}.

A single sweeping beam may create locations on a scattering surface where an actual pair of spots is created. Such a ``real" spot pair is defined as occurring when two places on the scattering surface become illuminated simultaneously in the inertial frame of the scattering surface. At a real spot pair creation location, the real spot speed on the scattering surface $v$ will formally diverge. These locations can be found by local extrema in the time of illumination from the beam relative to an arbitrary temporal zero point. There are surely many geometric situations that lead to such a speed divergence, but only two general scenarios are discussed below: scattering by a sphere and by a plane. 

\begin{figure}[h]
\includegraphics[width=18cm]{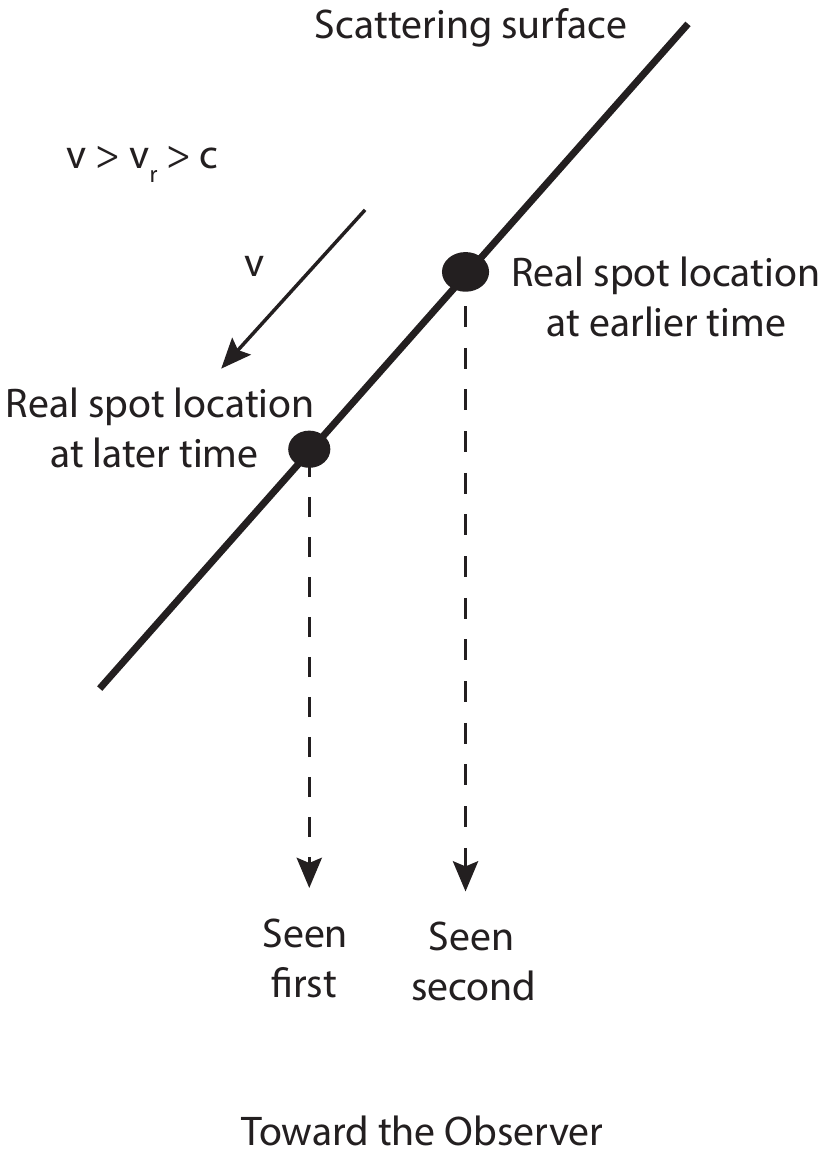}
\caption{A real spot is depicted moving superluminally along a scattering surface with $v_r > c$. Two locations of the spot are shown. Although the real spot is moving toward the lower left, the virtual image of the spot appears to the observer to be moving ``backward" toward the upper right.}
\label{TwoSpots}
\end{figure}

A simple but interesting case occurs when a spot moves across a scattering surface such that its speed toward the observer always exceeds the speed of light: $v_r > c$. In this case, parts of the scattering surface that are actually illuminated earlier will appear to the observer to be illuminated later. This is a simple kinematic effect -- for a given point on the scatterer, the path along the scatterer and then toward the observer has the first part moving superluminally toward the observer and the second part at $c$ toward the observer. In comparison, light taking the path from the given point directly to the observer always moves at speed $c$, and is therefore observed later. The situation is depicted in Figure \ref{TwoSpots}. One result is that superluminal spots with $v_r > c$ toward the observer will always appear to move away from the observer. Furthermore, were information coded temporally in these spots, that information would appear in the opposite time order to the observer than it sent from the beam source. 

When the projected speed toward the observer of a real spot goes from above $c$ to below $c$, the observer will always perceive a virtual spot pair creation event to occur at the $v_r = c$ location. To see this, consider a location on the scattering surface just a bit nearer to the observer than the $v_r = c$ location. Here, by definition, $v_r < c$. Light will necessarily take longer to reach the observer from this location than from the $v = c$ location because of the relative (subluminal) slowness of the spot on the surface. Therefore, the spot is seen first at the $v = c$ location.

Next consider a location on the scattering surface just a bit further from the $v_r = c$ location. Here, by definition, $v_r > c$. Light will also necessarily take longer to reach the observer from this location than from the $v = c$ location because of the extra distance it needs to travel to reach the observer. Therefore, again, the spot is seen first at the $v = c$ location. This leads to the perceived ``backward" motion of the spot from the same geometry depicted in Figure \ref{TwoSpots}.

Combining these two parts, it is clear that of the three locations, the $v_r = c$ location is seen first by the observer. Since just after this, locations on both sides of $v_r = c$ become visible, it can be concluded that spots at these locations are both seen by the observer {\it after} $v_r$ dropped from superluminal to subluminal. This spot pair creation event is virtual in the sense that no real spot pair creation event occurs at the $v_r = c$ location. 

Whenever a virtual spot pair appears, the observer perceives one spot from this pair to move along the scatterer in the same direction that the real spot is moving, so that $u_\perp$ shares a component moving along the surface in the same direction as $v$, while the other spot from this pair is perceived by the observer to move in the opposite direction, with $u_\perp$ sharing a component moving along the surface in the opposite direction as $v$. The ``forward" moving spot is perceived initially to move at formally infinite transverse speed $u_\perp$, but drops in magnitude as it moves along. Also, the ``backward" moving spot starts at formally infinite $u_\perp$ but in the opposite direction. As indicated above and in Figure \ref{TwoSpots}, the backward moving spot appears time-reversed -- if the real spot contained a beamed video, for example, then the observer would see this video playing backwards on the virtual spot moving away from the observer. The forward moving virtual spot does {\it not} appear to be time-reversed to the observer. 

Analogous logic to that given above can be used to show that when $v_r$ increases from subluminal to superluminal, a pair of existing virtual spots is seen by the observer to be annihilated. Here the $v_r = c$ location on the scattering surface can then be shown to be a temporal maximum so that spots on either side are always seen at an earlier time. This spot pair creation event is also virtual in the sense that no real spot pair creation event occurs at the $v_r = c$ location.

\section{A Sweeping Beam Scattered from a Sphere}

\subsection{Sweeping Beam with Constant Angular Speed}

\begin{figure}[h]
\includegraphics[width=18cm]{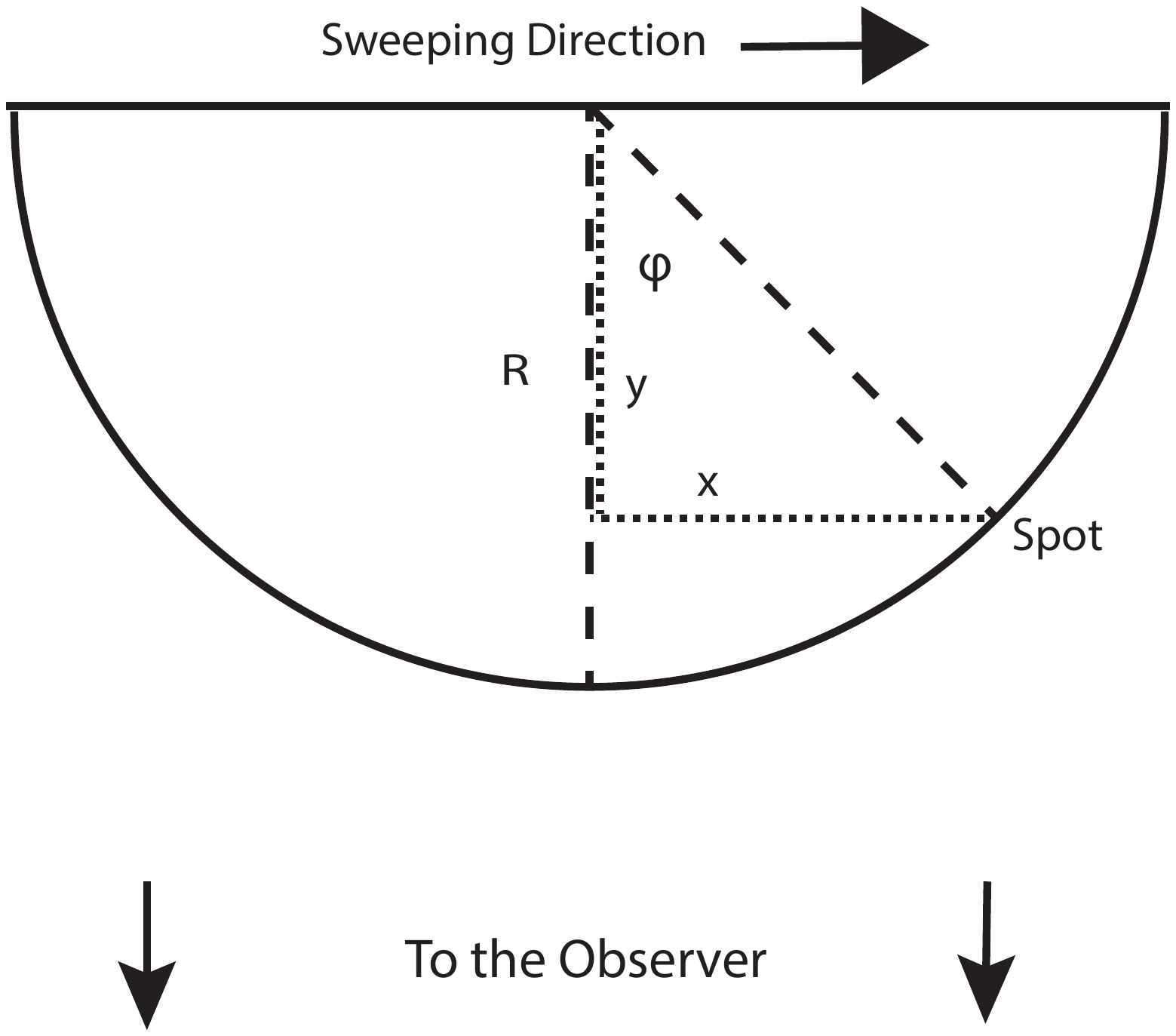}
\caption{The geometry of a sweeping beam that creates a spot or spots on a sphere.}
\label{GeometrySphere}
\end{figure}

The first canonical scenario involving superluminal spot pair events considered here will be that of a beam sweeping across a spherical body of radius $R$ at distance $D$ from an observer, with $R << D$, and with scattered light observed from very nearly the same direction as the outgoing beam. Assume that the angular size of the sphere is much larger than the angular size of the beamed spot on the sphere. Further assume that the beam sweeps linearly across the sphere once, through the central point of its projected disk, at a constant angular speed $\omega$ on the observer's sky. This projected hypothetical speed across a flattened disk at the distance of the sphere will be labelled $w_\perp = \omega D$. Let $\phi$ be the angle between the incoming beam and a given point on sphere in the beam illumination path, with the angle vertex being at the center of the sphere. Here $x$ labels the coordinate distance on the plane of the sky at distance $D$ from the observer along the sweeping beam, while $y$ labels a radial coordinate distance into the sky at distance $D$ from the observer. The origin of these $(x, y)$ coordinates is the center of the sphere, while $x = R \sin \phi$ and $y = R \cos \phi$. As the beam sweeps across the sphere, $\phi$ goes from $- \pi/2$ to $\pi/2$, $x$ goes from $-R$ to $R$, while $y$ goes from zero to $R$ and back to zero. The geometry is diagrammed in Figure \ref{GeometrySphere}. 

Considering the sphere a flat disk, the time it takes for the light beam to sweep across half the disk is $t_{sweep} =  R / w_\perp$. Assuming that sweeping begins pointing at $t_{delay} = 0$ toward $\phi = -\pi/2$ radians, then the time it takes for the light beam to point toward the location $\phi$ on the sphere can be quantified to be $t_{delay} = (R + x)/ w_\perp = R (1 + \sin \phi) / w_\perp$. 

Flying direct, the time it takes for light to cross half the sphere is $t = R / c$. The time it takes for light to go from the beam source to a location with coordinate $y$ on the sphere is $t_{path} = D / c + (R - y) / c$ where the first term is the time it takes for light to reach the closest point on the sphere, and the second term is the time it takes for light to cross distance $(R - y)$ of the sphere. Written in terms of $\phi$, then $t_{path} = D / c + R (1 - \cos \phi) / c$ and so the total time it takes before  position $\phi$ is illuminated will be  
 \begin{equation}
t_{real} = t_{delay} + t_{path} = R (1 + \sin \phi) / w_\perp + D / c + R (1 - \cos \phi) / c .
\end{equation}

Which part of the sphere is illuminated first? In general this is not $\phi = -\pi/2$ but rather is found from setting $d t_{real} / d {\phi} = 0$. The first illuminated point is therefore at $\phi_{first}^{real} = \arctan(-c / w_\perp)$. When $w_\perp >> c$ then $\phi_{first}^{real}$ goes to zero, the closest point on the sphere to beam source. When $w_\perp << c$ then $\phi_{first}$ goes to $-\pi/2$, the location on the sphere where the beam points first. When $w_\perp = c$, then $\phi_{first}^{real}$ goes to $-\pi/4$.  

Since every $\phi$ will be illuminated eventually, $\phi$ values on either side of $\phi_{first}^{real}$ are illuminated later, with pairs of $\phi$ locations being illuminated simultaneously. Therefore $\phi_{first}^{real}$ is also $\phi_{pair}^{real}$, and the first illuminated place on the sphere is actually a diverging pair of spots!

The real illumination pattern of a beam sweeping across a sphere can now be described. A real pair of beam spots is first illuminated at $\phi_{pair}^{real}$ with each spot moving in opposite directions. One spot moves toward the closest limb and disappears there, while the other spot crosses the rest of the sphere.

How fast do these spots move across the surface of the sphere?  The speeds are computed from 
\begin{equation} \label{vsphere}
v =  R {\ d\phi  \over  d t_{real} } = { w_\perp c \over c \ \cos \phi + w_\perp \sin \phi }.
\end{equation}
It is easily shown from the above Eq. (\ref{vsphere}) that the surface speed $v$ diverges at $\phi_{pair}^{real}$, with one spot moving out with initially infinite surface speed toward the $\phi = -\pi/2$ limb, with the other spot moving with initially infinite surface speed in the other direction -- toward $\phi = 0$.

To find the perceived illumination pattern by an observer, it will be useful to decompose $v$ into radial and perpendicular components. Then $v_\perp = v \cos \phi$ perpendicular to the observer and 
\begin{equation} \label{vparsphere}
v_r = - v \sin \phi = {- w_\perp \ c \ \sin \phi  \over
                                       c \ \cos \phi + w_\perp \ \sin \phi } ,
\end{equation}
radially toward the observer. Just as $v$ is unlimited in magnitude, the magnitude of $v_\perp$ and $v_r$ can exceed $c$. 

Starting from the time the light beam begins its journey at the source toward the sphere, to when the beam is measured back at the source by the observer, the time that angular position $\phi$ is observed to be illuminated is when 
 \begin{equation}
t_{obs} = t_{delay} + 2 t_{path} = R (1 + \sin \phi) / w_\perp + 2 D / c + 2 R (1 - \cos \phi) / c .
\end{equation}
Also $t_{obs} = t_{real} + t_{path}$. Which $\phi$ on the sphere is {\it observed} to be illuminated first? This will be when $d t_{obs} / d {\phi} = 0$, which occurs when $\phi_{pair}^{virtual} = \arctan (- c / 2 w_\perp)$.  The superscript ``virtual" highlights that no real spot pairs are created on the scatterer at this $\phi$ location. In general, the observed perpendicular speed of a spot will be 
\begin{equation} \label{uperpsphere}
u_\perp = R {d \phi \over d t_{obs} } = { w_\perp \ c \ \cos \phi \over 
                                                          c \ \cos \phi + 2 \ w_\perp \ \sin \phi } .
\end{equation}
Note that when $\phi = \phi_{pair}^{virtual}$, then $u_\perp$ formally diverges. Consequently, when $\phi$ is slightly less than $\phi_{pair}^{virtual}$ then $u_\perp$ has a very large negative value, meaning that the observed spot is initially moving very rapidly in the opposite direction than $w_\perp$. Alternatively, when $\phi$ is slightly greater than $\phi_{pair}^{virtual}$, then $u_\perp$ has a very large positive value, meaning that the observed spot is initially moving very rapidly in the same direction as $w_{\perp}$.

No matter how small the angular sweep speed across the observer's sky $\omega$, so long as it is finite, there is a location near the edge of the sphere where $v_r$ drops from greater than $c$ to less than $c$. Therefore, in all cases, an observer can perceive, in theory, a virtual spot pair creation event. This will also be the first light of any kind that an observer will see from the sweeping beam.

The angle $\phi_{pair}^{virtual}$ where a virtual pair of spots is first perceived is straightforward to compute. In Eq. (\ref{vparsphere}), $v_r$ is set equal to $c$. One then finds that $\phi_{pair}^{virtual} = \arctan (- c / 2 w_\perp)$, as indicated above. When the angular sweep speed $\omega$ is large, the arctangent goes to zero and therefore so does $\phi$, meaning that the spot pair creation event appears near the projected center of the sphere, the nearest point as seen by the observer. Conversely, when the sweep speed $\omega$ is low, then the arctangent goes to $-\pi / 2$ meaning that the virtual spot pair creation event is perceived to occur near the limb of sphere first pointed toward by the source.

The illumination pattern perceived by the observer of a beam sweeping across a sphere can now be adequately described. The very first thing the observer sees is a spot pair creation event with two spots simultaneously created at $\phi_{pair}^{virtual}$. One spot of this pair moves toward the nearby first-pointed-toward edge, counter-intuitively in the {\it opposite} direction from the actual motion of the sweeping beam. Simultaneously, a second spot moves toward the last-pointed-toward edge, in the {\it same} direction as the actual sweeping beam. The spot moving toward the first edge disappears at that limb before the spot moving toward the last edge. There is no spot pair annihilation event in this scenario.

A perhaps surprising feature is that one virtual spot appearing at $\phi_{pair}^{virtual}$ will subsequently be perceived to pass over $\phi_{pair}^{real}$, the $\phi$ location where real spot pair creation occurred, without anything unusual appearing to happen. Even though two real spots were created at $\phi_{pair}^{real}$, one of the virtual spots appears to move smoothly across. Therefore the {\it only} spot pair creation event witnessed by the observer is the one at $\phi_{pair}^{virtual}$. The observer sees nothing unusual happen at $\phi_{pair}^{real}$. 

Information about the angle of virtual spot pair creation, $\phi_{pair}^{virtual}$, is recoverable, theoretically, in at least three ways. The first detection method is by using both angular and temporal information -- by angularly resolving the spot pair creation event with sufficiently high speed imaging. The second method is purely temporal, by measuring the resulting light curve with sufficient detail. The third detection method utilizes polarization measurements of sufficiently high temporal sampling, discerning the changing polarization content of the scattered light.

\subsection{Sweeping Beams Across Spheres in Astronomical Settings}

As alluded to above, a popular example of a beam sweeping a spot across a sphere is a laser sweeping across Earth's Moon. The angular radius of the Moon is about 0.25 degrees, while the Moon's physical radius is about 1740 km. An easily noticeable spot pair creation event should occur when the linear sweep speed at the average distance of the Moon is $w_\perp = c \sim 300,000$ km $/$ sec. Were the Moon a flat disk, the time it would take for a laser to sweep across the Moon at this speed would be $t_{sweep} = 2 R_{Moon} / c \sim 0.0116$ seconds, which is just the light travel time across the Moon. At this rotation rate, a laser could sweep from one Earth horizon to the other, 180 degrees, in about 4.2 seconds. Creating spot pair events on the Moon with the average laser point is therefore simple and does not require expensive apparatuses. 

The relative brightness of spots created by a single beam sweep across the Moon is now estimated given three assumptions. The first assumption is that the Moon is a Lambertian reflector such that it returns the same brightness at all viewing angles, the second is that the beam size is large compared to surface scattering features such as craters and mountains but small compared to the Moon itself, and the third is that the beam sweeps with a constant angular speed on the observer's sky. Given these assumptions, the observed instantaneous brightness of a sweeping spot is proportional to $u_\perp$. The theory behind this simple relation starts by noting that each location along the swept path is both illuminated uniformly and scatters uniformly. Were the Moon a flat disk, it would just return a spot of unchanging brightness. The Moon's depth does {\it not} change the integrated brightness of each $\phi$ value. However, the depth of the truly three-dimensional Moon changes the timing and duration of when different $\phi$ values are illuminated and subsequently seen to be illuminated by the observer. Therefore, relatively, some illuminated swaths appear instantaneously bright for a short time, while others appear instantaneously dim for a long time.

\begin{figure}[h]
\includegraphics[angle=90, width=18cm]{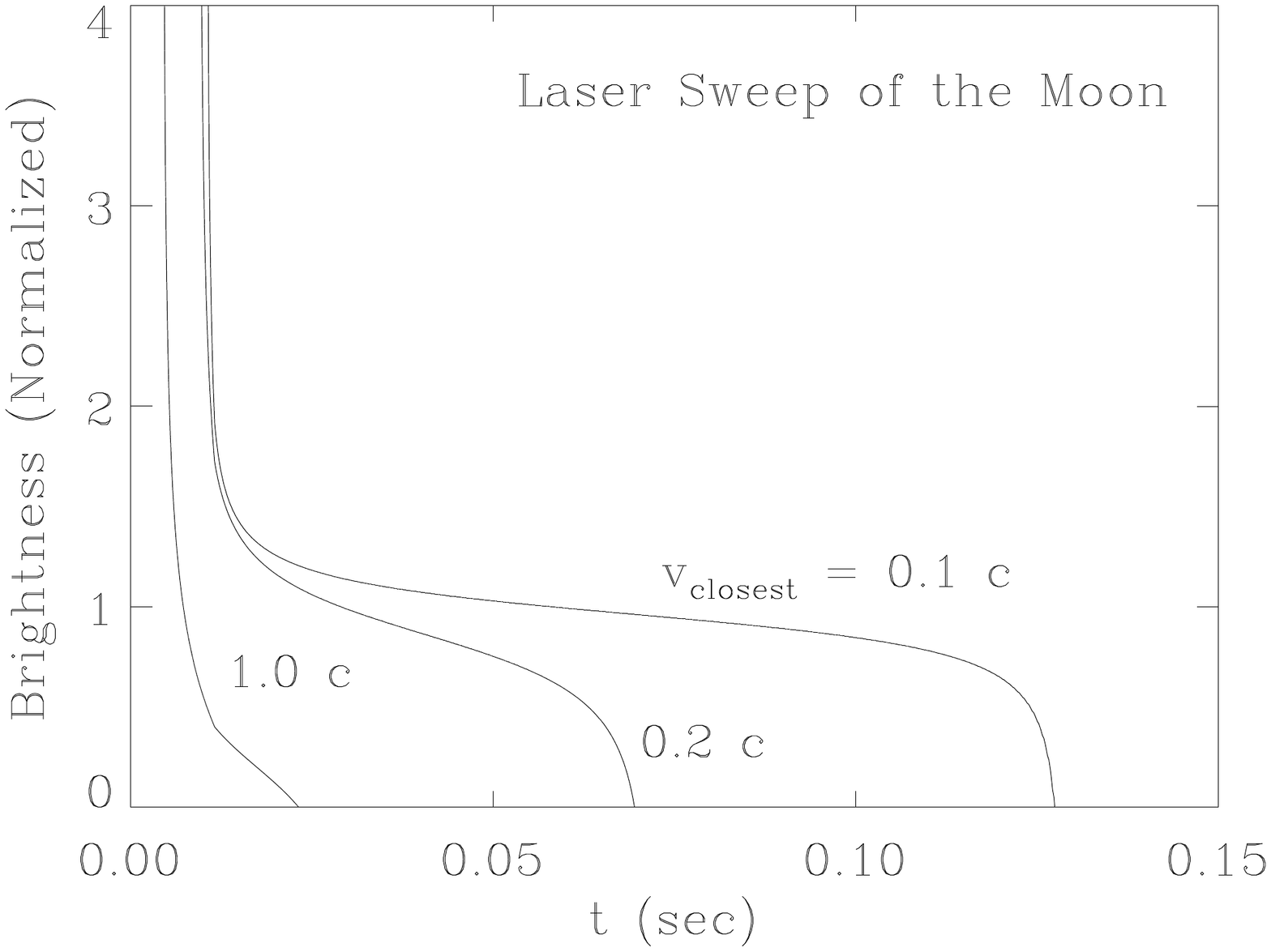}
\caption{A light curve of the instantaneous brightness of the spots created by a beam swept with constant angular speed across the Moon, as measured back on Earth. The curve labels refer to the spot sweep speed across the closest lunar point, where $\phi = 0$. The high initial brightnesses derive from perceived spot pair creation events being the first light that reaches the observer.}
\label{sweepsphere_bright3}
\end{figure}

Perceived instantaneous spot brightness on a sphere is depicted in Figure \ref{sweepsphere_bright3}, which plots this brightness as a function of time for a beam sweeping with the speeds $w_\perp =$ 0.1 c, 0.2 c, and 1.0 c respectively. Figure \ref{sweepsphere_bright3} was created under the assumption that no angular information is recovered and so gives the gross light curve measured instantaneously over the entire Moon. The faster sweeps show a higher early brightness -- formally infinite at $t = 0$ -- just as a superluminal virtual pair creation of spots is perceived. The slowest sweep speed $w_\perp = 0.1 c$ takes longer but still shows the initial virtual spot pair peak. The light curve then quickly flattens out to the brightness intuitively expected for a flat two-dimensional Moon, where $u_\perp = w_\perp$, which is also the normalized value. 

The formally infinite brightness appears because $u_\perp$ in Eq.(\ref{uperpsphere}) diverges for $\phi$ angles that make the denominator zero. In reality, the divergence would be muted by several factors including the infinitesimal amount of time that $u_\perp$ diverges, the finite size of the spot,  and the limited amount of energy emitted and scattered by the beam.

Unfortunately, as indicated in Figure \ref{sweepsphere_bright3}, the time scale for the virtual spot pair creation episode at the start of the light curves is a bit too brief to be discernable with the human eye. Still, effects of superluminal spot pair events should be discernable rapid imaging and a powerful laser sweeping past the lunar reflectors left by the NASA Apollo missions (Bender et al. 1973). Although these reflectors are too small to show a significant length of any sweeping beam, precisely kept times when specific discrete reflections are seen would test an underlying tenant of this analysis. 

A large spot boundary commonly observed to sweep across the Moon is the shadow of the Earth during a lunar eclipse. Given the above analysis, it should be clear that a lunar eclipse actually starts out as a virtual {\it pair} of dark shadow edges that suddenly appear very near a limb of the full Moon. One of these dark edges is perceived to move ``backwards" toward the closest limb and quickly disappears there, while the other appears to move progressively across the Moon as usually depicted. Towards the end of a lunar eclipse there occurs a creation event of a virtual pair of {\it bright} edges, again very near the Moon's limb. Again, one of the bright edges is perceived to move ``backwards" to the closest limb and very quickly disappears there, while the other edge appears to move progressively across the Moon as usually depicted. Unfortunately, all of this occurs within a small fraction of a second ($\sim 10^{-8}$ sec) after a given eclipse edge begins to cross the Moon and numerous effects including the fuzziness of Earth's shadow, due to the Earth's atmosphere, likely make the effect practically unobservable. In principle, this scenario works for other eclipse situations, for example eclipses of Jupiter by its moons. At Jupiter, the effect would last longer but still only be visible for only the order of microseconds.

A controlled sweeping beam could be used, in theory, to determine geometric surface characteristics of passing objects, including, for example, asteroids and comet nuclei, to determine how non-spherical they are. Most discerning would be a rapid series of sweeps, possibly in the radio or microwave bands, cycling through a range of orientations and stepping through an array of useful sweeping speeds. Each sweep in the series might itself be repeated numerous times to increase signal strength and to allow detection with a matched-frequency chopped or strobed detector. It may also be possible to change the shape and width of the sweeping beam to optimize sensitivity to surface characteristics slightly larger than the beam size. 

Regarding more distant astronomical settings, the beam of a pulsar may sweep a spot across the spherical surface of a companion star and hence create superluminal virtual spot pair creation events. As suggested by Milgrom \& Avni (1976) and further analyzed by Chester (1979), some of the X-rays from the binary pulsar 3U 0900-40 may be scattering off the surface of the primary companion and creating a signal possibly misinterpreted as orbital eccentricity. Recent theoretical work modeling this effect subluminally has been done by Dementyev (2014). 

\clearpage

\section{A Sweeping Beam Scattering from a Planar Wall or Linear Filament}

The second canonical case considered here involves superluminal spot pair events created by a beam sweeping across a planar reflecting wall. Since a sweeping beam itself defines a plane, and the intersection of two planes is a line, then the sweeping beam creates a straight line path on the wall. This case is conceptually equivalent to a large sweeping beam illuminating a smaller linear filament. For simplicity, unless stated otherwise, it will be assumed that the plane of the sweeping beam is perpendicular to the wall.

\subsection{Sweeping Beam with Constant Linear Speed}

\begin{figure}[h]
\includegraphics[width=18cm]{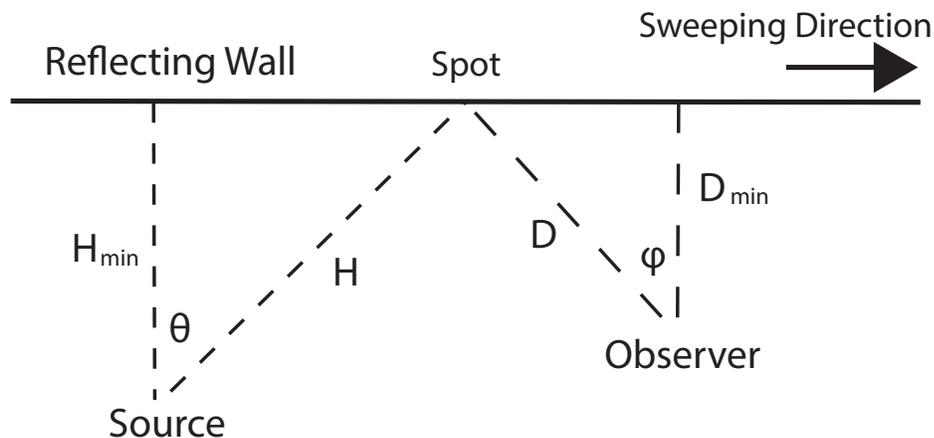}
\caption{The geometry of a sweeping beam that creates a spot or spots on a planar wall.}
\label{GeometryWall}
\end{figure}

A simple but informative scenario is that of a single spot moving with a constant linear speed $v$ across a scattering wall. The distance between the observer and a given point on the line of illumination will be labelled $D$ with minimum distance $D_{min}$. It will be assumed that $v > c$ so that superluminal spot pair effects can demonstrated. The direction with a component toward the observer will be considered the positive $v$ direction. The angle between the closest position on the sweeping beam line to the observer and the position of the beam on the line will be designated $\phi$, and the beam will be defined as moving from $\phi = -\pi/2$ to $\pi/2 $. The geometry is diagrammed on the right of Figure \ref{GeometryWall}.

The illuminated spot will start its motion at the $\phi = -\pi/2$ infinitely distant end of the swept line. The spot will then move along a line with a radial component toward the observer, pass the $\phi = 0$ point closest to the observer, and then move toward the $\phi = +\pi/2$ infinitely distant end of the line. 

\begin{figure}[h]
\includegraphics[width=18cm]{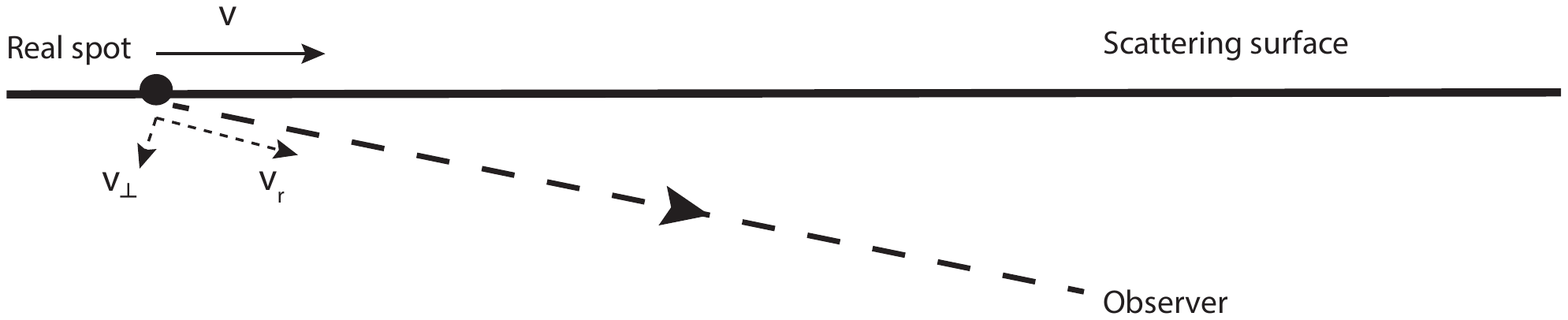}
\caption{The geometry of a spot moving with constant linear speed when the spot is far from the observer. Note that in this situation, most of the spot's speed $v$ is radially toward the observer, so that $v \sim v_r$.}
\label{vrfar}
\end{figure}

Initially, almost the entire spot velocity is directly radially toward the observer, so $v_r \sim v$. This is depicted in Figure \ref{vrfar}. Given that $v > c$, then $v_r > c$ at the start, but $v_r$ will decrease monotonically with increasing $\phi$ such that $v_r = - v \sin \phi$. Clearly, $v_r = 0$ when $\phi = 0$. Therefore, at some point, $v_r$ must cross from being greater than $c$ to being less than $c$, passing a location where $v_r = c$. This location is defined by $\phi_{pair}^{virtual} = {\rm arcsin} (-c/v)$. When $v >> c$, then $\phi_{pair}^{virtual}$ goes toward zero, the closest point on the sweeping beam line to the observer. 

Speed $v_r$ becomes equal to $c$ only once on this line. When the spot on the filament is at its closest to the observer, at $\phi = 0$, then by definition all of its speed is tangential, so that $v = v_\perp$ and $v_r = 0$. After the spot has passed $\phi = 0$, its projected speed is away from the observer and so $v_r < 0$. For the rest of this spot's trip, $\phi$ will be greater than zero, and $v_r < 0$ since the spot is headed away from the observer. In fact, for all positive and higher $\phi$ values, the spot's radial speed will always be negative. At some point $v_r$ will drop to below $-c$, but nothing unusual will be seen by the observer at the $v_r = -c$ location. As the spot finally approaches the end of the infinite filament, its speed is directly entirely away from the observer, so that $v_r \sim -v$.

To better quantify what an observer would see, it is useful to find the light travel times to the observer from different locations on the spot's path. The time it takes for a photon to go from a point on the spot's path to the observer will be designated $t_{path}$ and is equal to the path length of the light divided by $c$. At angle $\phi$, the distance between the observer and the spot is $D = D_{min} / \cos \phi$. Therefore $t_{path} = D / c = D_{min} / (c \ \cos \phi)$. 

Next, define the delay time $t_{delay}$ as the time between when the spot starts down the filament and the time when the filament element at angle $\phi$ is illuminated. Distance from the closest point to the observer to the point being considered along the filamentary line can be parameterized as $D_{filament} = \sqrt{ D^2 - D_{min}^2 } = D_{min} \ \tan \phi $. It is further assumed that the filament has total length $L$ which may be infinite. Then, $t_{delay} = L/2v - D_{filament} / v = L / 2 v + D_{min} \ \tan \phi / v$, where a negative $\phi$ indicates the spot is seen during approach.  

From the time the spot started down the filament to reach $\phi$, to the time light reaches the observer from the spot at $\phi$ is $t_{obs} = t_{delay} + t_{path}$ such that   
\begin{equation}
 t_{obs} = {L \over 2 v} + { D_{min} \over c \ \cos \phi}  
                   + {D_{min} \ \sin \phi \over v \ \cos \phi} .
 \end{equation}

The observer will first see the spot at the $\phi$ location where the total time it takes for light to reach the observer, $t_{obs}$, is at a minimum. Mathematically, this occurs when $d t_{obs} / d \phi = 0$. Since $v > c$, this does {\it not} occur infinitely far up the filament, and since it is a temporal minimum, spots will be perceived by the observer on both sides of this location at future times. Therefore the $\phi$ value at this location will be referred to as $\phi_{pair}^{virtual}$. One spot of the pair appears to the observer to move along the filament with a component toward the observer, while the other appears to move in the opposite direction. These two spots will appear to diverge to opposite ends of the filament. 

The transverse speed of the spot will be observed to be 
\begin{equation}
u_\perp = D {d \phi \over dt_{obs}} =
       { c \ v \cos \phi \over v \ \sin \phi + c} .
\end{equation}
This transverse speed diverges when $\phi = \phi_{pair}^{virtual}$. Specifically, when $\phi$ is slightly less than $\phi_{pair}^{virtual}$ then $u_\perp$ is negative and very large,  meaning that one image of the spot is seen to start its motion away from the observer quite quickly. Also, when $\phi$ is slightly greater than $\phi_{pair}^{virtual}$, $u_\perp$ is very large and positive, meaning that a second image of the same spot also appears to start its motion quite quickly, but in this case {\it toward} the observer.

In sum, even though only a single superluminal spot ever existed on the wall, the first thing the observer sees is a spot pair creation event at $\phi_{pair}^{virtual}$. The two perceived spots move away from each other, each, at first, with infinite angular speeds, but each quickly slowing. These two virtual spots will always remain visible to the observer, each always moving toward opposite ends of the filament. Note that when $v < c$ then $\phi_{pair}^{virtual}$ is not defined, meaning that subluminal real spots are never seen to create virtual pairs. In the above $v > c$ case, there is never any real spot pair creation event -- the existence of virtual spot pairs in this case is purely perceptual. 

\subsection{Sweeping Beam with Constant Angular Speed}

Another useful example occurs when a beam sweeps across a planar scattering wall at a constant {\it angular} speed, here parameterized as $\omega$. As before, the beam swept line on the wall is conceptually similar to an illuminated filament. The source of this spinning beam is considered at rest with respect to the scattering wall and the observer. With respect to the beam source, distances along the filament are given by the parameter $H$, with $H_{min}$ being the closest point on the filament to the source. With a vertex at the beam source, angles on the filament are labeled with the parameter $\theta$, with the furthest point on the filament in the initial direction of the beam to be $\theta = -\pi / 2$, the closest point to the source as $\theta=0$ and the furthest point opposite the initial direction of the beam to have $\theta = \pi/2$. Note that $\omega = {\dot \theta}$. The geometry is shown diagrammatically on the left part of Figure \ref{GeometryWall}.

This scenario starts with the light beam pointing parallel to the wall. As the beam tilts toward the wall, the first illuminated part of the wall will {\it not} be infinitely far from the source, at $\theta = -\pi/2$, because it will take an infinite time for light to reach that far from the beam source. Starting from the time when the beam points toward $\theta = -\pi/2$, the delay time it will take for the rotating beam to point toward position $\theta$ on the filament will be $t_{delay} = (\pi/2 + \theta) / \omega$. The time it takes for light to travel from the source to position $\theta$ on the filament will be the path length divided by the speed of light, so that $t_{path} = H_{min} / (c \ \cos \theta$).  From the start, the time that a filament position at $\theta$ will be illuminated (and so host a ``real" spot) will be 
\begin{equation} \label{treal}
t_{real} = t_{delay} + t_{path} = { \pi / 2 + \theta \over \omega} + {H_{min} \over c \ \cos \theta } .
\end{equation}
Now the first illuminated point of the filament occurs when $d t_{real} / d \theta = 0$, which occurs when $1/\omega + (H_{min} / c) (\sin \theta / \cos^2 \theta) = 0$. Therefore the $\theta$ of first illumination occurs when
\begin{equation}
{ \sin \theta \over \cos^2 \theta} = {- c   \over H_{min} \ \omega } ,
 \end{equation} 
which has solution
\begin{equation}
 \theta_{pair}^{real} = \arcsin \left( {H_{min} \omega \over 2 c} - \sqrt{ {H_{min}^2 \omega^2 \over 4 c^2} + 1 }  \right) .
\end{equation}

Since $\theta$ locations on either side of $\theta_{pair}^{real}$ will be illuminated {\it after} $\theta_{pair}^{real}$, a spot on one side of $\theta_{pair}^{real}$ will become illuminated at the same time as a spot on the other side. For this reason, $\theta_{pair}^{real}$ is considered the location of the creation of a real pair of spots. Interrogation of the $t_{real}$ Eq. (\ref{treal}) above indicates that the further that $\theta$ is from $\theta_{pair}^{real}$, the later in time it becomes illuminated. Therefore, the spots created at $\theta_{pair}^{real}$ will move on the filament away from $\theta_{pair}^{real}$ and each other. The analogous quantity to $\theta_{pair}^{real}$ in the previous section on beam-illuminated spheres is $\phi_{pair}^{real}$.

The speed of the beamed real spot on the filament perpendicular to the direction to the source is $H d \theta / dt_{real}$. Therefore the speed of the beamed real spot on the scattering wall is $v = (H / \cos \theta) d \theta / dt_{real}$ which gives 
 \begin{equation}
v = { c \ H_{min} \ \omega  \over c \ \cos^2 \theta + \omega \ H_{min} \  \sin \theta } .
\end{equation}

When $\theta = \theta_{pair}^{real}$, $v$ diverges. Also, $v$ changes sign on either side of $\theta_{pair}^{real}$, meaning that each real spot in the created pair moves in opposite directions on the filament, as indicated above. Even if $v$ at $\theta = 0$ is subluminal, $v$ may exceed $c$ at other values of $\theta$.

\begin{figure}[h]
\includegraphics[angle=90, width=18cm]{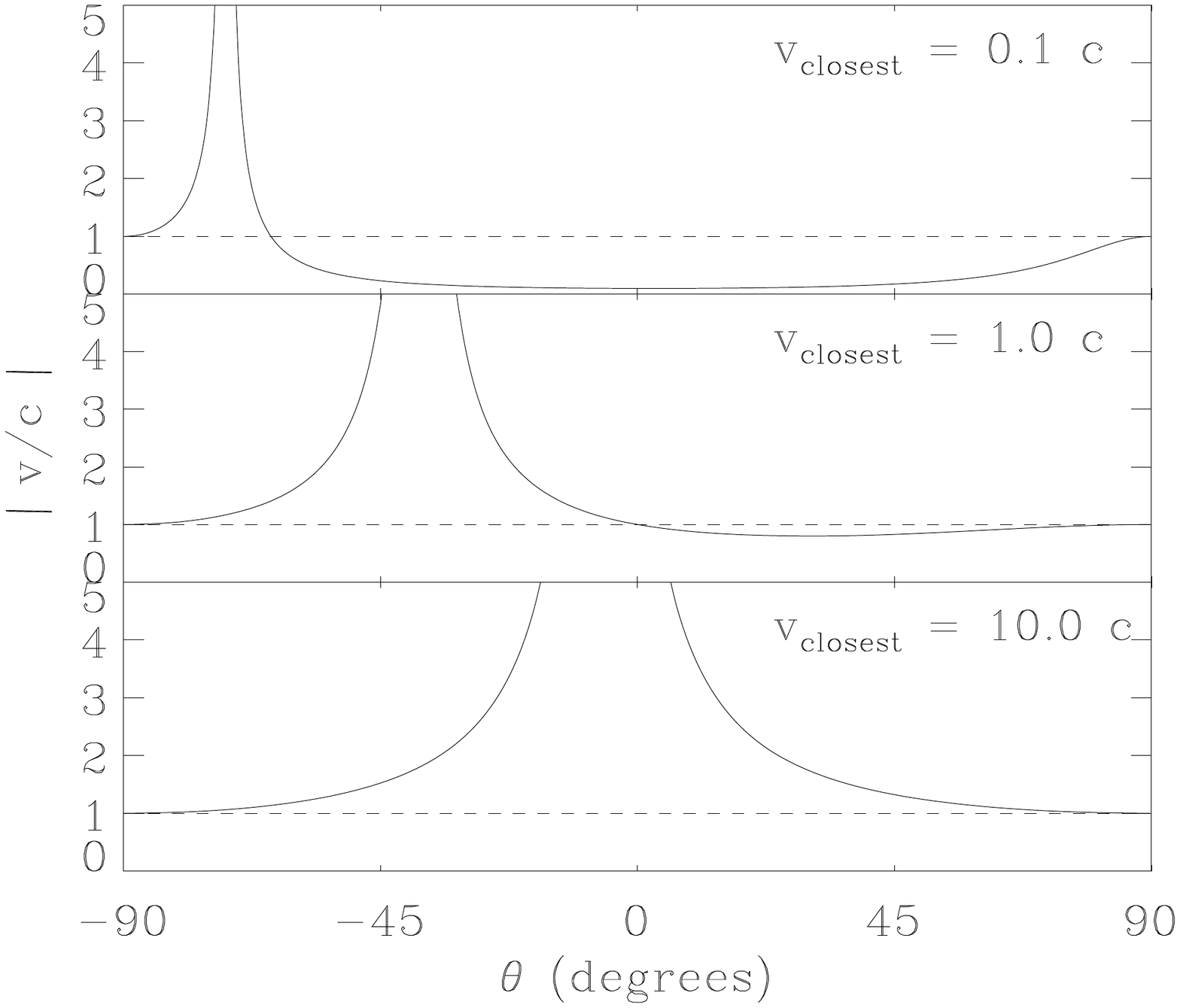}
\caption{The absolute value of the speed of real spots moving across a planar wall or linear filament is plotted against the beaming angle, for the case when the spots are created by a single fixed beam rotating with a constant angular speed in a plane perpendicular to the wall. The beam first points toward $\theta = -90^o$, moves to point toward the closest point on the scatterer at $\theta = 0^o$, and ends at $\theta = 90^o$. A divergent spike results from a real spot pair creation event and occurs for a beam with any finite angular speed. The superluminal spots mark the first section of the wall actually illuminated by the beam. The plot labels refer to the real spot speed at $\theta = 0$.}
\label{sweepwall_v3}
\end{figure}

A plot of the absolute value of real spot speed on the wall as a function of $\theta$ is shown in Figure \ref{sweepwall_v3} for three values of the speed across the closest section. Formally, $v_{closest} = \omega \ D_{min}$. In general, the lower the sweep speed, the closer the spot pair creation event will be to $\theta = -90$ degrees $= -\pi / 2$ radians. Conversely, the higher the sweep speed, the closer the real spot pair creation event will be to $\theta = 0$. The two lines on each plot depict the speed of each spot during a single sweep of constant angular speed. The real spot with the most negative $\theta$ will move in the opposite direction to that of the sweeping beam. This real spot is created at $\theta_{pair}^{real}$ with formally infinite speed and will always drop toward $\mid v \mid = c$ as $\theta$ drops to $-\pi/2$ radians, formally reaching $\mid v \mid = c$ at $\theta = - \pi/2$.

The real spot with the larger $\theta$ will move in the same direction as the sweeping beam and will also be created at $\theta_{pair}^{real}$ at formally infinite speed and at the same time as the other spot. Although this real spot will at first have its speed drop below $\mid v \mid = c$ as $\theta$ further increases, its speed will rise toward $\mid v \mid = c$ as $\theta$ rises toward $\pi/2$ radians, formally reaching $\mid v \mid = c$ at $\theta = \pi / 2$. 

What does an observer see? For the didactic purpose of enhancing the prevalence of superluminal spot pair effects, the observer is considered to be closest to the later part of the beam sweep, at positive $\theta$, as depicted in Figure \ref{GeometryWall}. As seen by the observer, angular placement of the illuminated sections of the filament will be labelled $\phi$ with $\phi$ starting at $- \pi/2$ and extending to $+\pi /2$. The distance between the observer and position $\phi$ on the filament is labelled $D$. The minimum distance between the observer and the illuminated filament on reflecting wall is labelled $D_{min}$ which occurs at $\phi = 0$. The spot's velocity along the filament can be broken up into components perpendicular and radial to the observer such that 
\begin{equation}
v_\perp = { c \ H_{min} \ \omega \  \cos \phi  \over c \ \cos^2 \theta + \omega \ H_{min} \  \sin \theta } ,
\end{equation}
and
\begin{equation}
v_r = { c \ H_{min} \ \omega \  \sin \phi  \over c \ \cos^2 \theta 
+ \omega \ H_{min} \  \sin \theta } .
\end{equation}
Note that $\theta$ and $\phi$ are not independent. Given positions of the source and observer, one can compute $\phi$ given $\theta$ and vice versa. Also, when $\phi$ is near $-\pi/2$ or $+\pi/2$ then $\theta$ will approach the same value, and vice versa.

To decipher what an observer would see, it is important to first find how $v_r$ changes as the beam sweeps through one cycle. First considering when $\phi$ is near $-\pi/2$, $v_r$ will be negative, meaning that a spot is perceived moving away from the observer. As $\phi$ increases, there will be discontinuous jump in $v_r$ where $v_r$ suddenly goes from negative infinity to positive infinity. For yet larger $\phi$ values, $v_r$ is positive, as a spot is moving toward the observer, but decreasing in magnitude. Eventually, as a spot's $v_r$ decreases, it will drop from being superluminal to subluminal. This $v_r = c$ location will be referred to as the spot ``virtual pair creation" location: $\phi_{pair}^{virtual}$. As $\phi$ increases further, $v_r$ will continue to decrease to zero and then past zero into negativity. Since as $\phi$ continues to increase, $v_r$ will only become more negative, then never again will $v_r$ drop from (positive) superluminal to subluminal. Therefore, no more virtual spot pair events will be observed. As $\phi$ approaches $\pi/2$, $v_r$ will continue to drop asymptotically toward $-c$ as the illuminating beam becomes increasingly parallel to the planar scattering wall. 

To better quantify what the observer will see, the timing of arriving photons will be calculated. Starting from the time when the source first started releasing photons as it pointed toward $\theta = -\pi/2$, the time it takes for a photon to reach the observer is
\begin{equation}
 t_{obs} = t_{delay} + t_{path}^{illum} + t_{path}^{obs} = t_{real} + t_{path}^{obs} 
            = { \pi / 2 + \theta \over \omega} + {H_{min} \over c \ \cos \theta } 
            + { D_{min} \over c \ \cos \phi} ,
\end{equation}
where $t_{delay}$ is the time it takes, since the start, for the beam to point at position $\theta$, $t_{path}^{illum}$ is the time it takes for light to go from the beam source to illuminate the scattering wall at position $\theta$, and $t_{path}^{obs}$ is the time it takes for light to go from position $\phi$ to the observer. The location of the observed (and hence virtual) spot pair event will be the $\phi_{pair}^{virtual}$ angle that satisfies $d t_{obs} / d \phi = 0$. Detailed inspection of how $t_{obs}$ changes with $\phi$ reveals what the observer will see and when. 

\begin{figure}[h]
\includegraphics[angle=90, width=18cm]{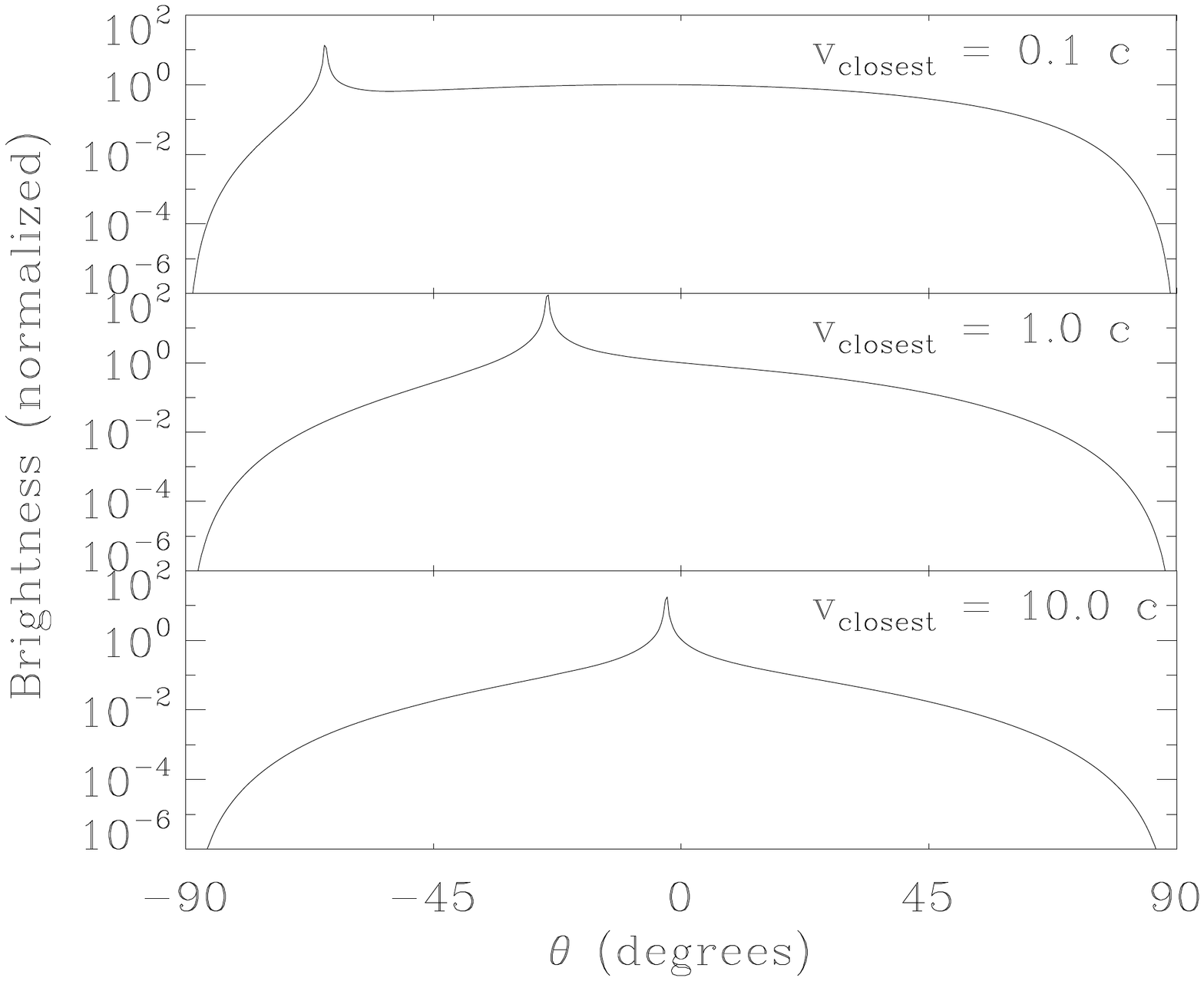}
\caption{The instantaneous perceived brightness of virtual spots perceived moving across a planar wall or linear filament, as observed from the beam source, is plotted against the beaming angle. Here the spots are created by a single fixed beam rotating with a constant angular speed in a plane perpendicular to the wall. The beam first points toward $\theta = -90^o$, moves to point toward the closest point on the scatterer at $\theta = 0^o$, and continues on to 90$^o$. The divergent spike results from a virtual spot pair creation event and occurs for a beam with any finite angular speed. The virtual spot pair creation event is the first light seen by the observer. The plot labels refer to the spot speed at $\theta = 0$.}
\label{sweepwall_amp3}
\end{figure}

Assuming that the wall is a Lambertian scatterer, it is straightforward to compute the relative brightness changes of the virtual spots visible to the observer. For increased simplicity, it will be further assumed here that the observer is at the source so that $\theta = \phi$. Specifically, using reasoning similar to the above lunar scenario, the instantaneous perceived brightness of a sweeping virtual spot as a function of $\theta$ is proportional to $b \propto u_\perp (D_{min}/D)^2$. This instantaneous perceived brightness is depicted in Figure \ref{sweepwall_amp3}. In this Figure, the unusual peak in brightness is caused by the perceived virtual spot pair creation event at $\phi_{pair}^{virtual}$. The high instantaneous perceived brightness is essentially caused by the relatively short time scale during which a relatively large part of the (nearly) uniformly bright scattering surface is scattering back light.
 
Given all of this detail, what an observer sees in this rotating beam scenario is surprisingly simple. The first phenomenon observed is a spot pair creation event at $\phi_{pair}^{virtual}$. The two virtual spots appear to move away from each other, each, at first, with infinite angular speeds, but each quickly slowing. These two virtual spots will always remain visible to the observer, each always moving toward the opposite distant ends of the wall or filament, and fading. This case is descriptively similar, as seen by the observer, to the previous case where the real spot speed was constant.

\subsection{Sweeping Beams Across Walls or Filaments in Astronomical Settings}

Similar but more complex scenarios than those considered above include eclipse light boundaries moving across the surface of reflection nebulae. Such mechanisms are thought to be the root cause of variable nebulae. Possibly the most notable variable nebula is Hubble's Variable Nebula (HVN: NGC 2261). The HVN, first noted by Hubble (1916), lies at a distance of about 2500 light years, estimated by an assumed association to the nearby open cluster NGC 2264 (Jones and Herbig 1982). Changes in the nebula's brightness have been attributed to shadows of opaque clouds moving between the bright star R Monocerotis and a reflection nebula (Johnson 1966, Lightfoot 1989). Seeming shifts in angular structure on the order of 0.5 arcminutes have been recorded over the time scale of tens of days. If attributable to single occulting events, these shifts indicate spot motions on the order of one light year per year, equivalent to $c$. This speed is an estimate of $v_\perp$ and not $v$ or $v_r$ and so does not directly reveal the key $v_r$ parameter that determines the perceived spot pair creation and annihilation events. Nevertheless, it seems reasonable to assume that given a long enough reflection train, virtual spot pair creation or annihilation events do occur as described above in the planar reflecting model. Additionally, sloping or bumpy terrain on the reflection nebula could well give rise to one or more $v_r = c$ crossings, and so yield virtual spot pair creation or annihilation events similar to that described in the spherical reflecting model. 

Although observers may be unlikely to see a spot pair creation event on the HVN without prior warning, pairs of spots moving away (toward) from each other might be observable from which a virtual spot pair creation (annihilation) event might be inferred. So long as the geometry of the scattering surface and the direction to the beam source remains relatively unchanged, the location of one spot pair event may also be the location of future spot pair events. For example, eclipses may show both ingress and egress events, and a single cloud moving near the source star might have multiple areas of high and low opacity. 

Other variable nebulae with potentially similar geometries that might show superluminal spot pair events include Hines Variable Nebula (see, for example, Moore 2005), Gyulbudaghian's Variable Nebula (the variable nebula associated with the variable star PV Cephei; Boyd 2012), infrared variable nebula IN L483 (Connelley, Hodapp, \& Fuller 2009), and NGC 6729 (the rapidly variable nebula associated with the star R Coronae Australis; see, for example Graham \& Phillips 1987). Additionally, the system UW Cen is a candidate for observerd virtual spot creation and annihilation events as it features an R Coronae Borealis star hypothesized to act as a lighthouse shining through gaps in thick dust clouds near its surface, illuminating changing portions of a surrounding reflection nebula (Clayton 2005). 

High frequency monitoring of variable nebulae might be able to find locations where virtual spot pair creation and annihilation events are occurring, and use these to build up information about the geometry of the surrounding reflecting surfaces. It is beyond the scope of this work to model these nebulae in detail but rather to raise the possibility that such effects might be occurring, discoverable in practice, and could yield information about the nebulae. A more specific investigation will be given in Zhong and Nemiroff (2015).  

Besides variable nebulae, another astronomical system that might show superluminal spot pair events are planetary nebulae. In particular, knots of optically thick dust in planetary nebulae are thought to cast shadows from the central star creating regions where specific ionizations do not occur. These shadows may move quickly as ionization fronts cross background gas, and could, in theory, move superluminally. Prominent possibilities include the NGC 7293 the Helix Nebula (O'Dell, Henney \& Ferland 2007), NGC 6543 the Cat's Eye Nebula (Balick 2004), and bi-polar planetary nebula M2-9 (van den Bergh 1974; Trammell, Goodrich, Dinerstein 1995; Corradi, Balick, \& Santander-Garc{\'{\i}}a 2011).

Another system type that might show superluminal spot pair events are circumstellar disks. Specifically, a bright star could create a silhouette of an opaque disk onto more distant reflecting material, enabling a shadow magnified in angular size by as much as a factor of 100 (Pontoppidan \& Dullemond 2005). Potentially, rapidly moving but subluminal inhomogeneities in the interior circumstellar disk could cast shadows moving superluminally onto a background. One example system is the Serpens Reflection Nebula and Ced 110 IRS 4 in the Chamaeleon I molecular cloud (Pontoppidan \& Dullemond 2005). Another is the case of HH 30 where an inner circumstellar disk is casting a large and variable shadow on an outer disk (Watson \& Stapelfeldt 2007). 

Pulsars surrounded by ionized shells provide yet another type of candidate system to cast superluminal shadows. In the radio, unusual ``moving" pulses from the Crab pulsar have been seen during several epochs and interpreted to be reflections of the beam off of ionized shell(s) in the outer part of the nebula (Lyne, Pritchard, Graham-Smith 2001).

\clearpage
\section{Discussion and Conclusions}

Given present knowledge of the geometry of several astronomical settings, it seems virtually certain that superluminal spots, shadows, and spot pair events do occur out in the universe. Possible venues include the Moon, nearby planets, passing comets and asteroids, variable nebula, pulsar jets, and jets in Herbig-Haro objects. What is less certain is whether virtual spot pair phenomena can be found in practical observing programs from Earth, and whether they can, in practice, reveal useful information. 

If found, superluminal virtual spot pair events could provide information that, theoretically, is not available from observations of subluminal spots: the radial real spot speed $v_r = c$. Given an independent measurement of the spot's transverse speed, this new radial component could yield an angular tilt measurement of the scatterer at the virtual spot pair event location. Conversely, modelling spot pair motion and flux changes may yield a good estimate of $v_\perp$, the true perpendicular velocity of the real spot near the virtual spot pair event location. Additionally, $u_\perp$ could be measured directly by a series of consecutive images. Since $v_\perp = D \ u_\perp$, comparing modelled $v_\perp$ to observed $u_\perp$ values may lead to an independent distance estimate $D$ to the scattering surface. Given the observed angle between the surface and the source, further constraints on the relative distances and angular speeds of opaque occulting clouds moving near bright sources might be recoverable. 

Sweeping beams are not confined to optical light, and in some systems other wavelength bands might yield more easily discernable virtual spot pair events. For passing asteroids, for example, beams in the radio and microwave bands -- otherwise used for radar -- might be better utilized. Were beam sizes smaller than surface structures deployed, the identification of virtual spot pair creation sites would encode information about the shape and size of these surface structures. Beyond illuminating beam spots and shadows, were a bright source of ionizing radiation considered, superluminal pairs of ionization fronts might be observable.  

To date, no clear superluminal spot pair creation or annihilation event has ever been reported. One reason is that the entire phenomenon is virtually unknown. Another reason is that discovery typically requires repeated observations of angularly extended systems. Dedicated monitoring of candidate systems has been typically sparse, to date. Potentially, the ability to detect superluminal spot pair events could be an impetus to increase monitoring, particularly if these events could yield discerning cloud dynamics or independent distance estimates. Developing and future sky survey projects such as Pan-STARRS and LSST may record multiple images in due course from which differential comparisons could reveal superluminal spot pair events. 

If the observer is able to control and repeat the sweeping of superluminal beams across a local target, a three dimensional map of the target object could be made, in principle. This is because any part of an object that scatters light can be swept with a beam multiple times, with multiple speeds, and at multiple angles until that part of the object shows superluminal spot pair events. Given a projected angular sweeping speed $w_\perp$ in the sweeping direction and that $v_r = c$ at the spot pair creation location, one can solve for the tilt $\phi_{pair}^{virtual}$ that this part of the scattering surface must have relative to the plane of the sky. The azimuthal direction of the tilt at this deflector location can be found by noting the sweep direction with the slowest $w_\perp$ that returns spot pair events. This whole process can then be repeated, in principle, for every observable parcel of the target object. The entire procedure can be considered a kind of superluminal pair spot tomography. 

A simple example of the potential utility of superluminal spot detection is in learning attributes of a sphere. Sweeping the sphere across its center at a given angular speed $\omega$ will generate a spot pair creation event at a $\phi = \phi_{pair}^{virtual}$ that can be input to Eq. (\ref{vsphere}) to find $w_\perp$, hence determining the distance $D$ to the sphere from only angular information. Conversely, $D$ can even be recovered by noting the shape of the light curve, and hence using only temporal information. Imagine now that this sphere itself has a small spherical bump on it -- the location and size of this bump can also be found by angular or temporal analyses of superluminal sweeps. Sweeping the sphere with a one-dimensional line (of spots) may even return all of this information without even knowing, at first, the precise angular location of the sphere.

The recoverable information from superluminal spot pair event analyses is different from the strictly angular information that occurs for single exposure photography and the strictly depth information that is obtained by single-illumination radar. Although the indicated use of an optical laser may indicate to some that this method is a type of lidar, the continuous sweeping beam and the reliance on superluminal spot pair events makes this method significantly different than standard timing-differential lidar.

It is of interest to recognize that due to superluminal spot pair events, the kinematics and observed motions of scattered spots from sweeping beams are, in general, {\it not} time symmetric. Consider, for example, a real superluminal spot moving along a wall. Only when moving toward the observer -- when its radial speed drops from superluminal to subluminal -- will an observable spot pair creation event occur. The same real spot does not create a virtual spot pair event after it passes the observer. Therefore, given a movie of a spot moving on a wall, one can use the virtual spot pair creation event to discern if the movie is being shown time-forward or time-backward. The creation and observation of superluminal spots moving on walls are therefore some of the more simple physical systems that shows a clear direction of time.

Sweeping beams are not the only mechanism that can create superluminal spot pair events. Another mechanism is the reflection of a spherically expanding flash of light off existing material.

Acknowledgements: The author acknowledges helpful conversations with Alexander Kostinski, Raymond Shaw, John Wallin, and Qi Zhong.


\begin{thebibliography}{}
%
\bibitem[Balick(2004)]{2004AJ....127.2262B} 
Balick, B.\ 2004, \aj, 127, 2262
%
\bibitem[Baune(2009)]{Baune2009}
Baune, S. \ 2009, Phys. Educ., 44, 296 
%
\bibitem[Bender et al.(1973)]{Bender1973} 
Bender, P.~L., Currie, D.~G., Dicke, R.~H., et al.\ 1973, Science, 182, 229
%
\bibitem[Blandford et al.(1977)]{Blandford1977} 
Blandford, R.~D., McKee, C.~F., \& Rees, M.~J.\ 1977, \nat, 267, 211
%
\bibitem[Boyd(2012)]{2012SASS...31...65B} 
Boyd, D.\ 2012, Society for Astronomical Sciences Annual Symposium, 31, 65
%
\bibitem[Cavaliere et al.(1971)]{Cavaliere1971} 
Cavaliere, A., Morrison, P., \& Sartori, L.\ 1971, Science, 173, 525
%
\bibitem[Chester(1979)]{1979ApJ...229.1085C} 
Chester, T.~J.\ 1979, \apj, 229, 1085
%
\bibitem[Clayton(2005)]{2005AIPC..804..180C} 
Clayton, G.~C.\ 2005, Planetary Nebulae as Astronomical Tools, 804, 180
%
\bibitem[Connelley et al.(2009)]{2009AJ....137.3494C} 
Connelley, M.~S., Hodapp, K.~W., \& Fuller, G.~A.\ 2009, \aj, 137, 3494
%
\bibitem[Corradi et al.(2011)]{2011A&A...529A..43C} 
Corradi, R.~L.~M., Balick, B., \& Santander-Garc{\'{\i}}a, M.\ 2011, \aap, 529, A43
%
\bibitem[Dementyev(2014)]{2014arXiv1403.4263D} 
Dementyev, A.~V.\ 2014, arXiv:1403.4263
%
\bibitem[Graham \& Phillips(1987)]{1987PASP...99...91G} 
Graham, J.~A., \& Phillips, A.~C.\ 1987, \pasp, 99, 91
%
\bibitem[Griffiths(1994)]{Griffiths1994}
Griffiths, D. J. 1994, Introduction to Quantum Mechanics (First Edition)
%
\bibitem[Hubble(1916)]{Hubble1916} 
Hubble, E.~P.\ 1916, \apj, 44, 190
%
\bibitem[Johnson(1966)]{Johnson1966} 
Johnson, H.~M.\ 1966, \aj, 71, 224
%
\bibitem[Jones \& Herbig(1982)]{Jones1982} 
Jones, B.~F., \& Herbig, G.~H.\ 1982, \aj, 87, 1223 
%
\bibitem[Lightfoot(1989)]{Lightfoot1989} 
Lightfoot, J.~F.\ 1989, \mnras, 239, 665
%
\bibitem[Lyne et al.(2001)]{2001MNRAS.321...67L} 
Lyne, A.~G., Pritchard, R.~S., \& Graham-Smith, F.\ 2001, \mnras, 321, 67
%
\bibitem[Milgrom \& Avni(1976)]{1976A&A....52..157M} 
Milgrom, M., \& Avni, Y.\ 1976, \aap, 52, 157
\bibitem[Moore(2005)]{2005JBAA..115...98M} 
Moore, S.\ 2005, Journal of the British Astronomical Association, 115, 98
%
\bibitem[Nemiroff(1993)]{Nemiroff1993}
Nemiroff, R. J. 1993, Am. J. Phys. 61, 619
%
\bibitem[O'Dell, Henney and Ferland(2007)]{ODell2007}
O'Dell, C. R., Henney, W. J. \& Ferland, G. J. 2007, ApJ 133, 2343
%
\bibitem[Pontoppidan \& Dullemond(2005)]{2005A&A...435..595P} 
Pontoppidan, K.~M., \& Dullemond, C.~P.\ 2005, \aap, 435, 595
%
\bibitem[Steane(2012)]{Steane2012}
Steane, Andrew 2012, The Wonderful World of Relativity: A Precise Guide for the General Reader. Oxford University Press. p. 180. ISBN 0-19-969461-3
%
\bibitem[Trammell et al.(1995)]{1995ApJ...453..761T} 
Trammell, S.~R., Goodrich, R.~W., \& Dinerstein, H.~L.\ 1995, \apj, 453, 761
%
\bibitem[van den Bergh(1974)]{1974A&A....32..351V} 
van den Bergh, S.\ 1974, \aap, 32, 351 

%
\bibitem[Watson  \& Stapelfeldt(2007)]{2007AJ....133..845W} 
Watson, A.~M., \& Stapelfeldt, K.~R.\ 2007, \aj, 133, 845
%
\bibitem[Zhong(2015)]{Zhong2015}
Zhong, Q. \& Nemiroff, R. J. 2015, in preparation
%
\end{thebibliography}
\end{document}